\documentclass[aps,prl,twocolumn,showpacs,superscriptaddress,letterpaper,amsmath,amssymb,nofootinbib]{revtex4-2}

\usepackage{graphicx}% Include figure files
\usepackage{dcolumn}% Align table columns on decimal point

\usepackage{amssymb}
\usepackage{amsmath}
\usepackage{gensymb}
\usepackage{bbm}
\usepackage{bm}
\usepackage{svg}
\usepackage{siunitx}

\usepackage{fancyhdr}
\usepackage{parskip}
\usepackage{dcolumn}
\newcommand{\galprop}{\mbox{GALPROP}}

\usepackage{mhchem}
\newcommand{\CTw}{$\ce{CO}$}
\newcommand{\CTh}{$\ce{^{13}CO}$}
\newcommand{\CEh}{$\ce{C^{18}O}$}

\newcommand{\hi}{$\ce{H\textsc{i}}$}
\newcommand{\hii}{$\ce{H\textsc{ii}}$}
\newcommand{\htwo}{$\ce{H_{2}}$}

\newcommand{\gray}{$\gamma$-ray}
\newcommand{\grays}{$\gamma$-rays}
\newcommand{\fermilat}{{\it Fermi}--LAT}

\begin{document}

\preprint{APS/PRL}

\title{Improved Modeling of the Discrete Component of the  Galactic Interstellar  \gray~Emission and Implications for the \textit{Fermi}--LAT Galactic Center Excess}

\author{Christopher M. Karwin}
\email{ckarwin@clemson.edu}
\affiliation{Department of Physics and Astronomy, Clemson University, Clemson, South Carolina 29634, USA}
\affiliation{Department of Physics and Astronomy, University of California, Irvine, California 92697, USA}
\author{Alex Broughton}
\email{abrought@uci.edu}
\affiliation{Department of Physics and Astronomy, University of California, Irvine, California 92697, USA}
\author{Simona Murgia}
\email{smurgia@uci.edu}
\affiliation{Department of Physics and Astronomy, University of California, Irvine, California 92697, USA}
\author{Alexander Shmakov}
\email{ashmakov@uci.edu}
\author{Mohammadamin Tavakoli}
\email{mohamadt@uci.edu}
\author{Pierre Baldi}
\email{pfbaldi@uci.edu}
\affiliation{Department of Computer Science, University of California, Irvine, California 92697, USA}

\date{\today}

\begin{abstract}
The aim of this work is to improve models for the \gray{} discrete or small-scale structure related to \htwo{} interstellar gas. Reliably identifying  this contribution is important to disentangle \gray{} point sources from interstellar gas, and to better characterize extended \gray{} signals. Notably, the \fermilat{} Galactic center (GC) excess, whose origin remains unclear, might be smooth or point-like. If the data contain a point-like contribution that is not adequately modeled, a smooth GC excess might be erroneously deemed to be point-like. We improve models for the \htwo{}-related \gray{} discrete emission for a 50$^\circ\times1^{\circ}$  region along the Galactic plane via \htwo{} proxies better suited to trace these features. We find that these are likely to contribute significantly to the \gray{} \fermilat{} data in this region, and the brightest ones are likely associated with detected \fermilat{} sources, a compelling validation of  this methodology. We discuss prospects to extend this methodology to other regions of the sky and implications for the characterization of the GC excess.  
\end{abstract}
\maketitle

The Galactic \gray~interstellar emission (IE) traces interactions of cosmic rays (CRs) with the interstellar medium, and constitutes most of the \gray~emission observed by \fermilat. Uncertainties in modeling the IE are large and difficult to constrain, and they impact the  study of other  \gray{} sources in the \fermilat{} data, point-like as well as extended. We focus on modeling  the small-scale  structure of the  IE which, if not robustly captured by the  model, confuses the determination of point sources, especially along the Galactic plane~\cite{Abdollahi_2020}. It was shown in~\cite{ajello2016fermi} that a large fraction of the point sources detected by \fermilat{} in the Galactic Center (GC) region could be misidentified  gas structure, and strongly dependent in number and spatial distribution on the IE model employed to analyze the data. This result indicates  the presence of point-like emission in the data arising from unmodeled structure in the interstellar gas, and it underlines   the importance of accurately  modeling this component  to reliably identify point sources  in the \gray~data.   

More reliable modeling of the small scale gas structure could also impact the  characterization of  extended sources. A prominent example is the \fermilat{} GC excess~\cite{goodenough2009,Hooper:2010mq,abazajian2011,abazajian2014,calore2014background,daylan2014characterization,ajello2016fermi}. Striking features of this excess are its spatial morphology and spectrum which are compatible  with annihilating dark matter (DM).  Alternative explanations have been proposed, with the leading one attributing the signal to a collection of discrete emitters such as an unresolved point source population of millisecond pulsars. The origin of the \textit{Fermi}-LAT GC excess remains a debated topic~\cite{doi:10.1146/annurev-nucl-101916-123029}. This debate could be settled by determining  whether the spatial morphology of the excess is consistent with a smooth distribution, as predicted for DM,  or with the cumulative  emission of a collection of point-like emitters. Statistical techniques have been employed to accomplish this~\cite{Bartels:2015aea,lee2016,Leane:2019xiy,Zhong:2019ycb}, including the non-Poissonian template fit (NPTF) which can detect upward fluctuations in the photon statistics above  Poisson noise which are associated with a collection of point sources. However, the results of the NPTF technique strongly depend on the modeling of the IE, and an uncontroversial resolution to the origin of the GC excess has not yet been reached. We posit that if the  interstellar gas is more structured and point-like  than current models predict,  the  unresolved point source contribution in the data could be erroneously inflated by the fainter component (also below detection threshold) of the  small scale gas features. In particular, statistical methods such as the NPTF might attribute this component to a smooth GC excess and conclude it is point-like.  These uncertainties might therefore  hinder the ability to distinguish  the smooth versus point-like duplicity of the excess.  We note that wavelet decomposition is another statistical technique that has been employed to resolve the GC excess. While these studies are not as impacted by the IE model, the related results on the nature of the GC excess remain uncertain nonetheless~\cite{Bartels:2015aea,2018PhRvD..98d3009B,2020PhRvL.124w1103Z}.  

In this work, we present a novel approach to improve modeling of the small-scale structure in the interstellar gas. 
The IE is  due primarily to CRs interacting with the interstellar hydrogen  gas (and radiation field),  in  molecular (\htwo), atomic (\hi), and ionized (\hii) forms. \htwo{} and \hi{} are  highly structured compared to \hii{}, and this structure  is traced by the related \gray{} emission. In this work, we focus on \htwo{} gas because of its high degree of structure and our methodology hinges on the availability of additional proxies better suited to  model it. The impact of the other components of the \gray{} IE (specifically \hi~and dark gas~\cite{2005Sci...307.1292G}), and related uncertainties, is not considered in this study and will be the focus of later work.  
Since \htwo{} does not emit at a characteristic radio frequency, other molecules are used to trace its distribution. In particular, carbon monoxide ($\mathrm{^{12}CO}$, or  \CTw ~hereafter)  is typically employed as a proxy. Radio surveys trace the distribution of \CTw{} across the sky and the \htwo{} column density is inferred by scaling the \CTw{} content by a conversion factor (${\rm X_{CO}}$) which gives the ratio of the integrated \CTw{} line emission to the \htwo{} column density. The bulk of the \htwo{} is traced following this methodology, and the survey of the \CTw{} $J$=1--0 transition line from \cite{2001ApJ...547..792D} has been widely employed. However, \CTw{} is typically optically thick in  the  denser cores of molecular clouds  and it  underestimates the total \htwo{} column density there. \gray{} IE models that employ \CTw{} to trace \htwo{} gas may therefore underestimate its finer structure. This limitation can be  addressed by exploiting  \CTw{} isotopologues, \CTh{} and \CEh,   also found in \htwo{} clouds. Although rarer, these isotopologues  are not as optically thick and therefore more reliable to probe  dense cores. 

We employ the data from the Mopra Southern Galactic Plane CO Survey (data release 3)~\cite{Braiding_2018} to trace the denser \htwo{} regions. The survey covers a 50 square degree region, spanning Galactic longitudes  $l$ = 300$^{\circ}$-350$^\circ$ and latitudes $|b| \le$ 0.5$^{\circ}$, and it targets the $J$ = 1--0 transitions of \CTw, \CTh, \CEh{}, and $\ce{C^{17}O}$. The data reduction process, which involves six main stages of processing to perform the band-pass correction and background subtraction,  yields data cubes of the brightness temperature for each spectral line, as a function of Galactic coordinates and local standard of rest velocity. Mopra's observations are carried out in   $1^{\circ}\times1^{\circ}$  segments, which we combine into a mosaic for the full region.  To ensure the highest quality data in the rare isotopologues, only pixels for which the brightness temperature exceeds the $1\sigma$ noise   provided by  Mopra are used. This step is necessary since the noise could otherwise predict spurious \gray~emission. In this analysis we utilize  the \CTw~and \CTh~data, as the \CEh{} emission is extremely sparse, and the $\ce{C^{17}O}$ emission was too faint to be detected in the survey. The spatial resolution is $0.6^{\prime}$ and the spectral resolution is 0.1~km~$\mathrm{sec^{-1}}$. For the region it surveys, the Mopra data provide a sharper view of the \CTw{} emission compared to \cite{2001ApJ...547..792D}, in addition to  probing the rarer \htwo{} tracers.

We calculate the  \htwo~column density  corresponding to Mopra's  \CTw{} and \CTh{}, referred to as    {$N(\mathrm{H_2})_{\mathrm{CO}}$} and {$ N(\mathrm{H_2})_{\mathrm{CO13}}$}, respectively. Following the method of~\cite{ackermann2012fermi}, the gas is  separated  into 17 Galactocentric radial bins based on its velocity and therefore corresponding to different distances from the GC, assuming uniform circular motion about the GC.  The $N(\mathrm{H_2})_{\mathrm{CO}}$  is determined as: 
\begin{equation}
\label{eq:hco}
    N(\mathrm{H_2})_{\mathrm{CO}} = W(\mathrm{CO})\times X_{\mathrm{CO}},  
\end{equation}
where $W(\mathrm{CO})$ is the line strength of the CO gas, and we adopt the radially-dependent $X_{\mathrm{CO}}$ from~\cite{ackermann2012fermi} for the 17 radial bins.

$N(\mathrm{H_2})_{\mathrm{CO13}}$ is evaluated as: \begin{equation}
\label{eq:hco13}
N(\mathrm{H_2})_{\mathrm{CO13}} = N(\mathrm{CO13}) \times \Big[\frac{\mathrm{H_2}}{\mathrm{CO13}}\Big].
\end{equation}
Following~\cite{2009tra..book.....W,Braiding_2018}, we derive the \CTh~column density $N(\mathrm{CO13})$ assuming that the gas is in local thermodynamic equilibrium at a fixed excitation temperature of 10 K.  For the abundance ratio $[\frac{\mathrm{H_2}}{\mathrm{CO13}}]$, we choose the upper bound of the range (2.7 - 7.5) $\times 10^5$ provided in~\cite{2008ApJ...679..481P}, in an attempt to assess the maximal  impact on the \gray~data, although it is not excluded that larger values are possible. 

The CR propagation code \galprop{} (v56) \cite{Moskalenko:1997gh,Moskalenko:1998gw,Strong:1998pw,Strong:1998fr,2006ApJ...642..902P,Strong:2007nh,Vladimirov:2010aq,Johannesson:2016rlh,porter2017high,Johannesson:2018bit,PhysRevC.98.034611} is used to evaluate the $\gamma$-ray intensity maps for the $\mathrm{H_2}$-related emission traced by    \CTw~and \CTh. \galprop{}   self-consistently calculates spectra and abundances of Galactic CR species and associated diffuse emissions ($\gamma$-rays, but also radio, X-rays) in 2D and 3D. 
We adopt the same \galprop{} input parameters as described in~\cite{karwin2019fermi}. 
GALPROP returns radially dependent  \gray~intensity all-sky maps  (in 17 Galactocentric annuli) which allow us to determine, for the region observed by Mopra,  the additional contribution in \grays~from the $\mathrm{H_2}$  traced by~\CTh, and assess its significance in simulated \fermilat~data. 
\begin{figure*}[t]
\centering
\includegraphics[width=1.0\textwidth]{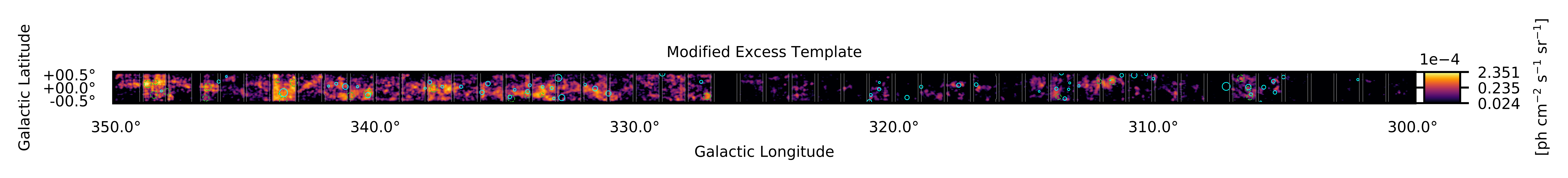}
\caption{The Modified Excess Template, calculated as the difference between the Modified Map and the baseline \CTw~map. The color scale shows the $\gamma$-ray intensity.  
The solid cyan circles show unassociated \gray{} point sources in version 4FGL-DR2  of the fourth \fermilat{} catalog~\cite{2020arXiv200511208B}, and the dashed green circles show new sources, not in the 4FGL-DR2,  that we find in this work (see text). The radius of each source corresponds to the 95\% localization uncertainty.}
\label{fig:excess_template}
\end{figure*}
The difference between {$ N(\mathrm{H_2})_{\mathrm{CO13}}$} and {$N(\mathrm{H_2})_{\mathrm{CO}}$} gives  an estimate of the \htwo~that is missed in dense regions when only CO is used as a tracer. To calculate the corresponding intensity  in \grays, we define the ``Modified Map". This is calculated in the following way: for pixels with  {$ N(\mathrm{H_2})_{\mathrm{CO13}}$}  $>$ {$N(\mathrm{H_2})_{\mathrm{CO12}}$}, the value of {$N(\mathrm{H_2})_{\mathrm{CO12}}$} is replaced with the value of {$ N(\mathrm{H_2})_{\mathrm{CO13}}$}.  We define the ``Modified Excess Template" as the difference between the Modified Map and the baseline \CTw~map,  
 which   accounts for the additional $\mathrm{H_2}$-related \gray{} emission not included in current IE models,  shown in Fig.~\ref{fig:excess_template}.

We evaluate the significance of the Modified Excess Template  in the \fermilat{} data  by simulating the data  collected  between 2008 August 04 to 2020 November  11 ($\sim$12 years). The simulated  events have energies in the range $1-100$ GeV and are binned in 8 energy bins per decade, for event class P8R3\_CLEAN (FRONT+BACK). The analysis is performed using Fermipy (v0.19.0), which utilizes the  Fermitools (v1.2.23). In these simulations, we only focus on the $\mathrm{H_2}$-related \gray~emission, and exclude all other components of the \gray~sky. The goal is to assess the  significance of the Modified Excess Template, i.e. the contribution of the newly modeled $\mathrm{H_2}$ fine structure, in the optimistic scenario where  all other components are satisfactorily  modeled. The  simulated events  trace the $\mathrm{H_2}$-related $\gamma$-ray emission modeled with the Modified Map. The simulated data are then fit based on a binned maximum likelihood method to a model that  includes two components, the  \grays~traced by the baseline \CTw~map and  the Modified Excess Template. The  latter  is assigned the spectrum determined by GALPROP, and its  normalization is  free to vary in the fit.  The normalization of the \CTw~map is also free to vary and its spectrum constrained to that calculated by GALPROP. As mentioned above, the \gray~flux is calculate  in 17 radial bins, since the predicted $\mathrm{H_2}$-related $\gamma$-ray emission depends on the CR density, which is a function of Galactocentric radius. In the simulations,  the total emission is integrated along the line-of-sight. Moreover, the individual maps have a high level of degeneracy. We therefore combine the maps into 4 radial bins. Specifically, we combine bins 1-6, 7-10, 11-13, and 14-17, which we refer to as A1, A2, A3, and A4.
 
\begin{figure}
\centering
\includegraphics[width=0.5\textwidth]{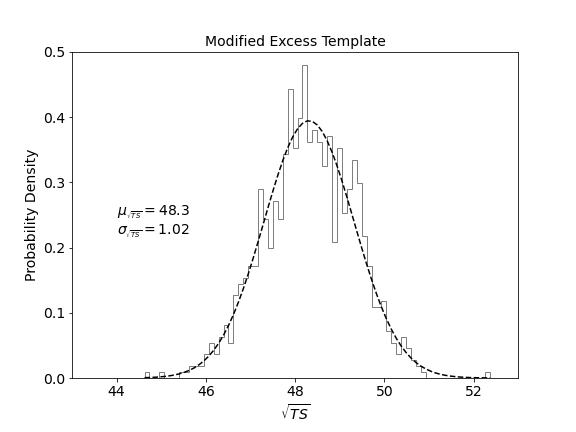}
\caption{Distribution of the statistical significance ($\sigma \approx \sqrt{\mathrm{TS}}$) of the Modified Excess Template for the 50$^{\circ}\times$1$^{\circ}$ region covered by Mopra for 1000 realizations of $\sim$12 years of \fermilat~ data. A fit with a Gaussian function is overlaid. }
\label{fig:excess_significance}
\end{figure}

We simulate 1000 realizations of \fermilat~data, and calculate the Test Statistics (TS) for the nested models ($\mathrm{-2log(L_0/L)}$, where $\mathrm{L_0}$ corresponds to value of the likelihood function for the null hypothesis (CO baseline), and $\mathrm{L}$ to the alternative hypothesis (CO baseline and Modified Excess Template.) The statistical significance is approximated by $\sigma \approx \sqrt{\mathrm{TS}}$.  The distribution of the $\sqrt{TS}$ for the 1000 simulations is shown in Fig.~\ref{fig:excess_significance}. The mean of the distribution is 48.30$\pm$1.02, for the 50 squared degree Mopra region, and therefore very significant in a  scenario where other components contributing to the \fermilat~data are perfectly modeled, and if the \gray~emission traced by \CTh~is at the high end of the range we have considered (with the caveats discussed above.) 

\begin{figure}
\centering
\includegraphics[width=0.5\textwidth]{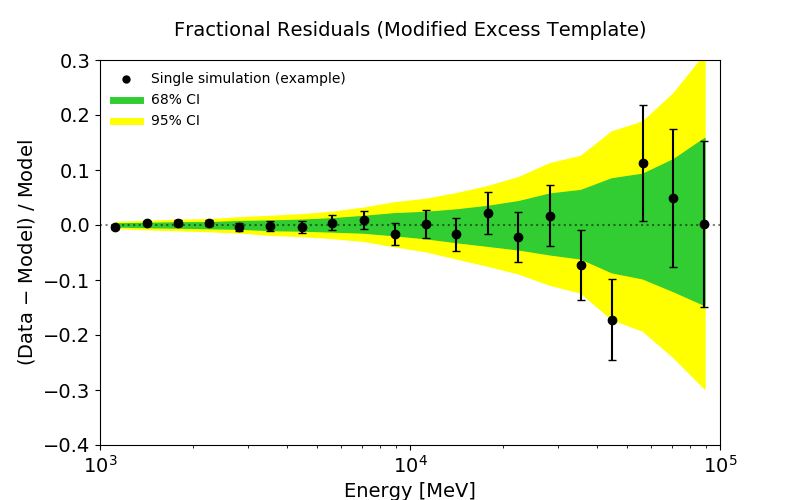}
\caption{Fractional count residuals as a function of energy. The green and yellow bands show the 68\% and 95\% confidence regions from 1000 simulations, respectively. As an example, we also plot the results for a single simulation, shown with black data points.}
\label{fig:gof}
\end{figure}

\begin{figure*}
\centering
\includegraphics[width=1.0\textwidth]{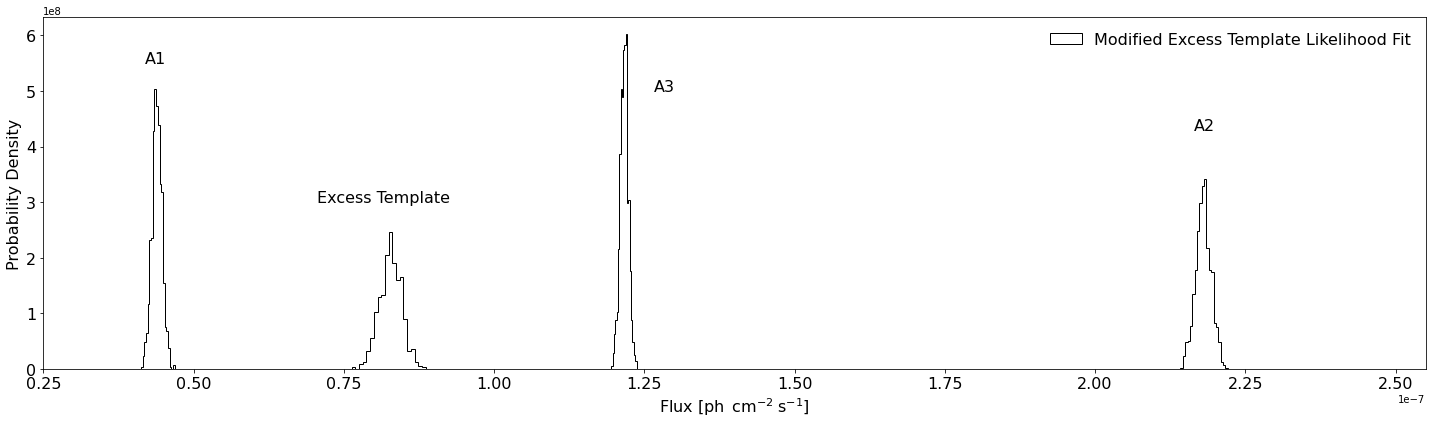}
\caption{The distributions of fluxes for each component of the best-fit model using the Modified Excess Template  over the full region for 1000 realizations of $\sim$ 12 years of \fermilat~data. Note that one of the annuli (A4)  has negligible contribution and the normalization was fixed in the fitting process (corresponding to a flux of $4.98 \times 10^{-9} $ \si{ph.cm^{-2}.s^{-1}}).}
\label{fig:main_hist2}
\end{figure*}

The fractional  residuals as a function of energy for the 1000 simulations are shown in Fig.~\ref{fig:gof}. They are consistent with zero, as expected.  
Fig.~\ref{fig:main_hist2} shows the distributions of the  flux of each model component, including the Modified Excess Template.  In  the Mopra region, the mean integrated flux is $(7.2 \pm 0.2) \times 10^{-8} \ \mathrm{ph \ cm^{-2} \ s^{-1}}$. 
Overall,  the Modified Excess Template accounts for a fair fraction (15.4\%) of the total \htwo-related \gray~emission in the Mopra region.  For comparison, the integrated GC excess  flux in a  15$^{\circ}\times$15$^{\circ}$   region around the GC, which is 4$\times$ larger than the region observed by Mopra, is in the range $18.3-25.0\times 10^{-8}$ $\mathrm{ph \ cm^{-2} \ s^{-1}}$ (from~\cite{ajello2016fermi}). In intensities, the GC excess  corresponds to $2.67 - 3.65 \times 10^{-6}~\mathrm{ph \ cm^{-2} \ s^{-1} \ sr^{-1}}$ compared to $4.72 \times 10^{-6}~\mathrm{ph \ cm^{-2} \ s^{-1} \ sr^{-1}}$ for the Modified Excess Template. Because of the different spatial morphology of the GC excess and \htwo{}-related \gray{} emission traced by \CTh{}, a direct comparison  is unwarranted and we do not expect the latter  to   account for the majority of the GC excess.  However, the $\mathrm{H_2}$ emission extends beyond the latitudes considered in this study and, albeit dimmer at higher latitudes, the estimates provided here do not indicate its small scale component  to be negligible. This contribution could therefore   confuse the GC excess morphology,  depending on the exact \htwo{} emission outside of $\pm$0.5$^{\circ}$ in latitude, because the techniques that track the point-like fluctuations could  erroneously ascribe  unaccounted point-like emission (originating from \htwo{} gas in this case) to the GC excess~\cite{Leane:2019xiy}. Considerations based on the spectrum of the unresolved source population could be  powerful to settle this degeneracy, however  a robust determination is required.

We identify the structure traced by \CTh{} that is bright enough to be detected as a \fermilat~point source and compare it to the unassociated  sources in version 4FGL-DR2  of the fourth \fermilat{} catalog of \gray{} sources~\cite{2020arXiv200511208B}. A significant overlap would indicate that its brightest contribution has already been detected by \fermilat~ and validate the methodology to trace this component. To this end, we perform a likelihood fit where, instead of including  the Modified Excess Template in the model, we only consider the baseline \CTw~map and find additional point sources using the fermipy \textit{find\_srcs} function. The method calculates TS maps, and identifies point sources based on peaks in the TS.  We use a test point source modeled with a power law spectrum with spectral index=2, and a minimum TS threshold of TS $\ge$ 16. An index of 2.7 is compatible with  CRs interacting with gas, and if  used, it yields consistent results.     
In total we find 23 new point sources with TS$\geq$16, which are shown with green circles in Fig.~\ref{fig:excess_template}. For comparison, we also overlay 4FGL-DR2 unassociated sources. We find that 8/23 ($35\%$) of the new sources are spatially consistent  with an unassociated 4FGL-DR2 source (based on an overlap of the 95\% localization errors), which accounts for 8/46 ($17\%$) of the total unassociated sources in the region. 
To quantify the probability that the associations happen by chance, we perform 1000 simulations, where for each realization we randomly distribute 23 sources in the Mopra region. For each source the Galactic coordinates are drawn from a uniform distribution and we assign an error radius drawn from a Gaussian distribution, with the mean and standard deviation determined from the 23 detected sources. From the 1000 simulations we find the average number of random associations to be 3.16 $\pm$ 1.72, which corresponds to a p-value (i.e.~N($\geq8$)/1000) of 0.012. It is therefore likely that the  brightest of the point-like emission traced by Mopra \CTh~have already been detected  in the \fermilat~data.

The results presented here demonstrate  that there likely is   significant     structure  in the \htwo-related  \gray~emission detected by \fermilat{} that is not currently included in the IE models. Since its  spatial morphology has point-like features, it  directly affects the detection of resolved and, potentially, unresolved point source populations in the \gray~data.   More specifically, we have discussed the impact on the 4FGL-DR2 catalog and on the interpretation of the GC excess. Although in this analysis we have focused on the region covered by Mopra,  similar conclusions would hold elsewhere because of the same limitations in tracing  $\mathrm{H_2}$.
The  methodology described here however can only  be readily applied to very small regions of the sky because  of the paucity of observations of the rare \htwo{} tracers  (e.g.~\cite{blackwell_burton_rowell_2016}).  Also, we cannot simply extrapolate the Modified  Excess Template with the less ambitious goal of providing only an estimate of this emission since  the  Mopra data on which it is based  is tightly confined, especially in latitude.  To address the limited available  observations, we have   resorted   to   machine learning to develop a methodology that  predicts the distribution of the small scale \htwo-related \gray~emission for other regions of the sky  based on the existing Mopra data. Because of its complexity, we describe the machine related work  in a  companion paper~\cite{companion}.

\textit{Conclusions}.--- We assess whether there is point-like emission in the~\gray~data associated with the interstellar gas and not currently included in models of the Galactic interstellar \gray~emission. We employ the  data from the Mopra Southern Galactic Plane CO Survey,  which includes tracers of the small scale structure of the $\mathrm{H_2}$-related \grays,  to   improve  models by more accurately describing the point-like features in the gas. We find  that significant point-like emission  originates from   $\mathrm{H_2}$ gas and the significance  could be as high as  $\sqrt{TS} \sim $ 48 (depending on assumptions) in a 50$^{\circ}\times$1$^{\circ}$ region covered by Mopra, corresponding  to $\sim$ 15\% of the modeled \htwo-related \gray~emission in the region.  We also show that this newly found point-like component may account for some fraction ($\lesssim$17\%) of  \gray~point sources detected by \fermilat~in the Galactic plane whose origin is so far unknown.    That a significant number of unidentified  4FGL-DR2 sources along the Galactic plane  originate from unmodelled structure in the gas is not unexpected, but here we develop a robust and reliable methodology to identify the contribution of both  bright and dimmer components of the \htwo~gas discrete emission. Given the significance of this emission, its contribution in the GC region could introduce a significant systematic uncertainty in  determining whether the GC excess is  smooth or point-like, and therefore more consistent with a dark matter or pulsar interpretation,   specifically when relying on statistical techniques  such as the NPTF.  Identifying  this component in the GC region could be a crucial  step to settle the origin  of the GC excess and to reliably determine whether  it is smooth and therefore a signal from DM. 

 \section{Acknowledgements}
We thank Troy Porter for many helpful discussions and insights. The work of AS, MT, and PB is in part supported by grants NSF NRT 1633631 and ARO  76649-CS to PB.

\bibliography{Bibliography}

\begin{thebibliography}{36}%
\makeatletter
\providecommand \@ifxundefined [1]{%
 \@ifx{#1\undefined}
}%
\providecommand \@ifnum [1]{%
 \ifnum #1\expandafter \@firstoftwo
 \else \expandafter \@secondoftwo
 \fi
}%
\providecommand \@ifx [1]{%
 \ifx #1\expandafter \@firstoftwo
 \else \expandafter \@secondoftwo
 \fi
}%
\providecommand \natexlab [1]{#1}%
\providecommand \enquote  [1]{``#1''}%
\providecommand \bibnamefont  [1]{#1}%
\providecommand \bibfnamefont [1]{#1}%
\providecommand \citenamefont [1]{#1}%
\providecommand \href@noop [0]{\@secondoftwo}%
\providecommand \href [0]{\begingroup \@sanitize@url \@href}%
\providecommand \@href[1]{\@@startlink{#1}\@@href}%
\providecommand \@@href[1]{\endgroup#1\@@endlink}%
\providecommand \@sanitize@url [0]{\catcode `\\12\catcode `\$12\catcode
  `\&12\catcode `\#12\catcode `\^12\catcode `\_12\catcode `\%12\relax}%
\providecommand \@@startlink[1]{}%
\providecommand \@@endlink[0]{}%
\providecommand \url  [0]{\begingroup\@sanitize@url \@url }%
\providecommand \@url [1]{\endgroup\@href {#1}{\urlprefix }}%
\providecommand \urlprefix  [0]{URL }%
\providecommand \Eprint [0]{\href }%
\providecommand \doibase [0]{https://doi.org/}%
\providecommand \selectlanguage [0]{\@gobble}%
\providecommand \bibinfo  [0]{\@secondoftwo}%
\providecommand \bibfield  [0]{\@secondoftwo}%
\providecommand \translation [1]{[#1]}%
\providecommand \BibitemOpen [0]{}%
\providecommand \bibitemStop [0]{}%
\providecommand \bibitemNoStop [0]{.\EOS\space}%
\providecommand \EOS [0]{\spacefactor3000\relax}%
\providecommand \BibitemShut  [1]{\csname bibitem#1\endcsname}%
\let\auto@bib@innerbib\@empty
%</preamble>
\bibitem [{\citenamefont {Abdollahi~{\it et al}}(2020)}]{Abdollahi_2020}%
  \BibitemOpen
  \bibfield  {author} {\bibinfo {author} {\bibfnamefont {S.}~\bibnamefont
  {Abdollahi~{\it et al}}},\ }\bibfield  {title} {\bibinfo {title} {Fermi large
  area telescope fourth source catalog},\ }\href
  {https://doi.org/10.3847/1538-4365/ab6bcb} {\bibfield  {journal} {\bibinfo
  {journal} {The Astrophysical Journal Supplement Series}\ }\textbf {\bibinfo
  {volume} {247}},\ \bibinfo {pages} {33} (\bibinfo {year} {2020})}\BibitemShut
  {NoStop}%
\bibitem [{\citenamefont {Ajello}\ \emph {et~al.}(2016)\citenamefont {Ajello},
  \citenamefont {Albert}, \citenamefont {Atwood}, \citenamefont {Barbiellini},
  \citenamefont {Bastieri}, \citenamefont {Bechtol}, \citenamefont
  {Bellazzini}, \citenamefont {Bissaldi}, \citenamefont {Blandford},
  \citenamefont {Bloom} \emph {et~al.}}]{ajello2016fermi}%
  \BibitemOpen
  \bibfield  {author} {\bibinfo {author} {\bibfnamefont {M.}~\bibnamefont
  {Ajello}}, \bibinfo {author} {\bibfnamefont {A.}~\bibnamefont {Albert}},
  \bibinfo {author} {\bibfnamefont {W.}~\bibnamefont {Atwood}}, \bibinfo
  {author} {\bibfnamefont {G.}~\bibnamefont {Barbiellini}}, \bibinfo {author}
  {\bibfnamefont {D.}~\bibnamefont {Bastieri}}, \bibinfo {author}
  {\bibfnamefont {K.}~\bibnamefont {Bechtol}}, \bibinfo {author} {\bibfnamefont
  {R.}~\bibnamefont {Bellazzini}}, \bibinfo {author} {\bibfnamefont
  {E.}~\bibnamefont {Bissaldi}}, \bibinfo {author} {\bibfnamefont
  {R.}~\bibnamefont {Blandford}}, \bibinfo {author} {\bibfnamefont
  {E.}~\bibnamefont {Bloom}}, \emph {et~al.},\ }\bibfield  {title} {\bibinfo
  {title} {Fermi-lat observations of high-energy $\gamma$-ray emission toward
  the galactic center},\ }\href@noop {} {\bibfield  {journal} {\bibinfo
  {journal} {The Astrophysical Journal}\ }\textbf {\bibinfo {volume} {819}},\
  \bibinfo {pages} {44} (\bibinfo {year} {2016})}\BibitemShut {NoStop}%
\bibitem [{\citenamefont {Goodenough}\ and\ \citenamefont
  {Hooper}(2009)}]{goodenough2009}%
  \BibitemOpen
  \bibfield  {author} {\bibinfo {author} {\bibfnamefont {L.}~\bibnamefont
  {Goodenough}}\ and\ \bibinfo {author} {\bibfnamefont {D.}~\bibnamefont
  {Hooper}},\ }\href {https://doi.org/10.48550/ARXIV.0910.2998} {\bibinfo
  {title} {Possible evidence for dark matter annihilation in the inner milky
  way from the fermi gamma ray space telescope}} (\bibinfo {year} {2009}),\
  \Eprint {https://arxiv.org/abs/0910.2998} {arXiv:0910.2998 [hep-ph]}
  \BibitemShut {NoStop}%
\bibitem [{\citenamefont {Hooper}\ and\ \citenamefont
  {Goodenough}(2011)}]{Hooper:2010mq}%
  \BibitemOpen
  \bibfield  {author} {\bibinfo {author} {\bibfnamefont {D.}~\bibnamefont
  {Hooper}}\ and\ \bibinfo {author} {\bibfnamefont {L.}~\bibnamefont
  {Goodenough}},\ }\bibfield  {title} {\bibinfo {title} {{Dark Matter
  Annihilation in The Galactic Center As Seen by the Fermi Gamma Ray Space
  Telescope}},\ }\href {https://doi.org/10.1016/j.physletb.2011.02.029}
  {\bibfield  {journal} {\bibinfo  {journal} {Phys. Lett. B}\ }\textbf
  {\bibinfo {volume} {697}},\ \bibinfo {pages} {412} (\bibinfo {year}
  {2011})},\ \Eprint {https://arxiv.org/abs/1010.2752} {arXiv:1010.2752
  [hep-ph]} \BibitemShut {NoStop}%
\bibitem [{\citenamefont {Abazajian}(2011)}]{abazajian2011}%
  \BibitemOpen
  \bibfield  {author} {\bibinfo {author} {\bibfnamefont {K.~N.}\ \bibnamefont
  {Abazajian}},\ }\bibfield  {title} {\bibinfo {title} {The consistency of
  fermi-{LAT} observations of the galactic center with a millisecond pulsar
  population in the central stellar cluster},\ }\href
  {https://doi.org/10.1088/1475-7516/2011/03/010} {\bibfield  {journal}
  {\bibinfo  {journal} {Journal of Cosmology and Astroparticle Physics}\
  }\textbf {\bibinfo {volume} {2011}}\bibinfo  {number} { (03)},\ \bibinfo
  {pages} {010}}\BibitemShut {NoStop}%
\bibitem [{\citenamefont {Abazajian}\ \emph {et~al.}(2014)\citenamefont
  {Abazajian}, \citenamefont {Canac}, \citenamefont {Horiuchi},\ and\
  \citenamefont {Kaplinghat}}]{abazajian2014}%
  \BibitemOpen
\bibfield  {number} {  }\bibfield  {author} {\bibinfo {author} {\bibfnamefont
  {K.}~\bibnamefont {Abazajian}}, \bibinfo {author} {\bibfnamefont
  {N.}~\bibnamefont {Canac}}, \bibinfo {author} {\bibfnamefont
  {S.}~\bibnamefont {Horiuchi}},\ and\ \bibinfo {author} {\bibfnamefont
  {M.}~\bibnamefont {Kaplinghat}},\ }\bibfield  {title} {\bibinfo {title}
  {Astrophysical and dark matter interpretations of extended gamma ray emission
  from the galactic center},\ }\href
  {https://doi.org/10.1103/PhysRevD.90.023526} {\bibfield  {journal} {\bibinfo
  {journal} {Physical Review D}\ }\textbf {\bibinfo {volume} {90}} (\bibinfo
  {year} {2014})}\BibitemShut {NoStop}%
\bibitem [{\citenamefont {Calore}\ \emph {et~al.}(2015)\citenamefont {Calore},
  \citenamefont {Cholis},\ and\ \citenamefont
  {Weniger}}]{calore2014background}%
  \BibitemOpen
  \bibfield  {author} {\bibinfo {author} {\bibfnamefont {F.}~\bibnamefont
  {Calore}}, \bibinfo {author} {\bibfnamefont {I.}~\bibnamefont {Cholis}},\
  and\ \bibinfo {author} {\bibfnamefont {C.}~\bibnamefont {Weniger}},\
  }\bibfield  {title} {\bibinfo {title} {Background model systematics for the
  fermi {GeV} excess},\ }\href {https://doi.org/10.1088/1475-7516/2015/03/038}
  {\bibfield  {journal} {\bibinfo  {journal} {Journal of Cosmology and
  Astroparticle Physics}\ }\textbf {\bibinfo {volume} {2015}}\bibinfo  {number}
  { (03)},\ \bibinfo {pages} {038}}\BibitemShut {NoStop}%
\bibitem [{\citenamefont {Daylan}\ \emph {et~al.}(2016)\citenamefont {Daylan},
  \citenamefont {Finkbeiner}, \citenamefont {Hooper}, \citenamefont {Linden},
  \citenamefont {Portillo}, \citenamefont {Rodd},\ and\ \citenamefont
  {Slatyer}}]{daylan2014characterization}%
  \BibitemOpen
\bibfield  {number} {  }\bibfield  {author} {\bibinfo {author} {\bibfnamefont
  {T.}~\bibnamefont {Daylan}}, \bibinfo {author} {\bibfnamefont {D.~P.}\
  \bibnamefont {Finkbeiner}}, \bibinfo {author} {\bibfnamefont
  {D.}~\bibnamefont {Hooper}}, \bibinfo {author} {\bibfnamefont
  {T.}~\bibnamefont {Linden}}, \bibinfo {author} {\bibfnamefont {S.~K.}\
  \bibnamefont {Portillo}}, \bibinfo {author} {\bibfnamefont {N.~L.}\
  \bibnamefont {Rodd}},\ and\ \bibinfo {author} {\bibfnamefont {T.~R.}\
  \bibnamefont {Slatyer}},\ }\bibfield  {title} {\bibinfo {title} {The
  characterization of the gamma-ray signal from the central milky way: A case
  for annihilating dark matter},\ }\href
  {https://doi.org/https://doi.org/10.1016/j.dark.2015.12.005} {\bibfield
  {journal} {\bibinfo  {journal} {Physics of the Dark Universe}\ }\textbf
  {\bibinfo {volume} {12}},\ \bibinfo {pages} {1} (\bibinfo {year}
  {2016})}\BibitemShut {NoStop}%
\bibitem [{\citenamefont
  {Murgia}(2020)}]{doi:10.1146/annurev-nucl-101916-123029}%
  \BibitemOpen
  \bibfield  {author} {\bibinfo {author} {\bibfnamefont {S.}~\bibnamefont
  {Murgia}},\ }\bibfield  {title} {\bibinfo {title} {The fermi-lat galactic
  center excess: Evidence of annihilating dark matter?},\ }\href
  {https://doi.org/10.1146/annurev-nucl-101916-123029} {\bibfield  {journal}
  {\bibinfo  {journal} {Annual Review of Nuclear and Particle Science}\
  }\textbf {\bibinfo {volume} {70}},\ \bibinfo {pages} {455} (\bibinfo {year}
  {2020})},\ \Eprint
  {https://arxiv.org/abs/https://doi.org/10.1146/annurev-nucl-101916-123029}
  {https://doi.org/10.1146/annurev-nucl-101916-123029} \BibitemShut {NoStop}%
\bibitem [{\citenamefont {Bartels}\ \emph {et~al.}(2016)\citenamefont
  {Bartels}, \citenamefont {Krishnamurthy},\ and\ \citenamefont
  {Weniger}}]{Bartels:2015aea}%
  \BibitemOpen
  \bibfield  {author} {\bibinfo {author} {\bibfnamefont {R.}~\bibnamefont
  {Bartels}}, \bibinfo {author} {\bibfnamefont {S.}~\bibnamefont
  {Krishnamurthy}},\ and\ \bibinfo {author} {\bibfnamefont {C.}~\bibnamefont
  {Weniger}},\ }\bibfield  {title} {\bibinfo {title} {{Strong support for the
  millisecond pulsar origin of the Galactic center GeV excess}},\ }\href
  {https://doi.org/10.1103/PhysRevLett.116.051102} {\bibfield  {journal}
  {\bibinfo  {journal} {Phys. Rev. Lett.}\ }\textbf {\bibinfo {volume} {116}},\
  \bibinfo {pages} {051102} (\bibinfo {year} {2016})},\ \Eprint
  {https://arxiv.org/abs/1506.05104} {arXiv:1506.05104 [astro-ph.HE]}
  \BibitemShut {NoStop}%
\bibitem [{\citenamefont {Lee}\ \emph {et~al.}(2016)\citenamefont {Lee},
  \citenamefont {Lisanti}, \citenamefont {Safdi}, \citenamefont {Slatyer},\
  and\ \citenamefont {Xue}}]{lee2016}%
  \BibitemOpen
  \bibfield  {author} {\bibinfo {author} {\bibfnamefont {S.~K.}\ \bibnamefont
  {Lee}}, \bibinfo {author} {\bibfnamefont {M.}~\bibnamefont {Lisanti}},
  \bibinfo {author} {\bibfnamefont {B.~R.}\ \bibnamefont {Safdi}}, \bibinfo
  {author} {\bibfnamefont {T.~R.}\ \bibnamefont {Slatyer}},\ and\ \bibinfo
  {author} {\bibfnamefont {W.}~\bibnamefont {Xue}},\ }\bibfield  {title}
  {\bibinfo {title} {Evidence for unresolved $\ensuremath{\gamma}$-ray point
  sources in the inner galaxy},\ }\href
  {https://doi.org/10.1103/PhysRevLett.116.051103} {\bibfield  {journal}
  {\bibinfo  {journal} {Phys. Rev. Lett.}\ }\textbf {\bibinfo {volume} {116}},\
  \bibinfo {pages} {051103} (\bibinfo {year} {2016})}\BibitemShut {NoStop}%
\bibitem [{\citenamefont {Leane}\ and\ \citenamefont
  {Slatyer}(2019)}]{Leane:2019xiy}%
  \BibitemOpen
  \bibfield  {author} {\bibinfo {author} {\bibfnamefont {R.~K.}\ \bibnamefont
  {Leane}}\ and\ \bibinfo {author} {\bibfnamefont {T.~R.}\ \bibnamefont
  {Slatyer}},\ }\bibfield  {title} {\bibinfo {title} {{Revival of the Dark
  Matter Hypothesis for the Galactic Center Gamma-Ray Excess}},\ }\href
  {https://doi.org/10.1103/PhysRevLett.123.241101} {\bibfield  {journal}
  {\bibinfo  {journal} {Phys. Rev. Lett.}\ }\textbf {\bibinfo {volume} {123}},\
  \bibinfo {pages} {241101} (\bibinfo {year} {2019})},\ \Eprint
  {https://arxiv.org/abs/1904.08430} {arXiv:1904.08430 [astro-ph.HE]}
  \BibitemShut {NoStop}%
\bibitem [{\citenamefont {Zhong}\ \emph {et~al.}(2020)\citenamefont {Zhong},
  \citenamefont {McDermott}, \citenamefont {Cholis},\ and\ \citenamefont
  {Fox}}]{Zhong:2019ycb}%
  \BibitemOpen
  \bibfield  {author} {\bibinfo {author} {\bibfnamefont {Y.-M.}\ \bibnamefont
  {Zhong}}, \bibinfo {author} {\bibfnamefont {S.~D.}\ \bibnamefont
  {McDermott}}, \bibinfo {author} {\bibfnamefont {I.}~\bibnamefont {Cholis}},\
  and\ \bibinfo {author} {\bibfnamefont {P.~J.}\ \bibnamefont {Fox}},\
  }\bibfield  {title} {\bibinfo {title} {{Testing the Sensitivity of the
  Galactic Center Excess to the Point Source Mask}},\ }\href
  {https://doi.org/10.1103/PhysRevLett.124.231103} {\bibfield  {journal}
  {\bibinfo  {journal} {Phys. Rev. Lett.}\ }\textbf {\bibinfo {volume} {124}},\
  \bibinfo {pages} {231103} (\bibinfo {year} {2020})},\ \Eprint
  {https://arxiv.org/abs/1911.12369} {arXiv:1911.12369 [astro-ph.HE]}
  \BibitemShut {NoStop}%
\bibitem [{\citenamefont {{Balaji}}\ \emph {et~al.}(2018)\citenamefont
  {{Balaji}}, \citenamefont {{Cholis}}, \citenamefont {{Fox}},\ and\
  \citenamefont {{McDermott}}}]{2018PhRvD..98d3009B}%
  \BibitemOpen
  \bibfield  {author} {\bibinfo {author} {\bibfnamefont {B.}~\bibnamefont
  {{Balaji}}}, \bibinfo {author} {\bibfnamefont {I.}~\bibnamefont {{Cholis}}},
  \bibinfo {author} {\bibfnamefont {P.~J.}\ \bibnamefont {{Fox}}},\ and\
  \bibinfo {author} {\bibfnamefont {S.~D.}\ \bibnamefont {{McDermott}}},\
  }\bibfield  {title} {\bibinfo {title} {{Analyzing the gamma-ray sky with
  wavelets}},\ }\href {https://doi.org/10.1103/PhysRevD.98.043009} {\bibfield
  {journal} {\bibinfo  {journal} {\prd}\ }\textbf {\bibinfo {volume} {98}},\
  \bibinfo {eid} {043009} (\bibinfo {year} {2018})},\ \Eprint
  {https://arxiv.org/abs/1803.01952} {arXiv:1803.01952 [astro-ph.HE]}
  \BibitemShut {NoStop}%
\bibitem [{\citenamefont {{Zhong}}\ \emph {et~al.}(2020)\citenamefont
  {{Zhong}}, \citenamefont {{McDermott}}, \citenamefont {{Cholis}},\ and\
  \citenamefont {{Fox}}}]{2020PhRvL.124w1103Z}%
  \BibitemOpen
  \bibfield  {author} {\bibinfo {author} {\bibfnamefont {Y.-M.}\ \bibnamefont
  {{Zhong}}}, \bibinfo {author} {\bibfnamefont {S.~D.}\ \bibnamefont
  {{McDermott}}}, \bibinfo {author} {\bibfnamefont {I.}~\bibnamefont
  {{Cholis}}},\ and\ \bibinfo {author} {\bibfnamefont {P.~J.}\ \bibnamefont
  {{Fox}}},\ }\bibfield  {title} {\bibinfo {title} {{Testing the Sensitivity of
  the Galactic Center Excess to the Point Source Mask}},\ }\href
  {https://doi.org/10.1103/PhysRevLett.124.231103} {\bibfield  {journal}
  {\bibinfo  {journal} {\prl}\ }\textbf {\bibinfo {volume} {124}},\ \bibinfo
  {eid} {231103} (\bibinfo {year} {2020})},\ \Eprint
  {https://arxiv.org/abs/1911.12369} {arXiv:1911.12369 [astro-ph.HE]}
  \BibitemShut {NoStop}%
\bibitem [{\citenamefont {{Grenier}}\ \emph {et~al.}(2005)\citenamefont
  {{Grenier}}, \citenamefont {{Casandjian}},\ and\ \citenamefont
  {{Terrier}}}]{2005Sci...307.1292G}%
  \BibitemOpen
  \bibfield  {author} {\bibinfo {author} {\bibfnamefont {I.~A.}\ \bibnamefont
  {{Grenier}}}, \bibinfo {author} {\bibfnamefont {J.-M.}\ \bibnamefont
  {{Casandjian}}},\ and\ \bibinfo {author} {\bibfnamefont {R.}~\bibnamefont
  {{Terrier}}},\ }\bibfield  {title} {\bibinfo {title} {{Unveiling Extensive
  Clouds of Dark Gas in the Solar Neighborhood}},\ }\href
  {https://doi.org/10.1126/science.1106924} {\bibfield  {journal} {\bibinfo
  {journal} {Science}\ }\textbf {\bibinfo {volume} {307}},\ \bibinfo {pages}
  {1292} (\bibinfo {year} {2005})}\BibitemShut {NoStop}%
\bibitem [{\citenamefont {{Dame}}\ \emph {et~al.}(2001)\citenamefont {{Dame}},
  \citenamefont {{Hartmann}},\ and\ \citenamefont
  {{Thaddeus}}}]{2001ApJ...547..792D}%
  \BibitemOpen
  \bibfield  {author} {\bibinfo {author} {\bibfnamefont {T.~M.}\ \bibnamefont
  {{Dame}}}, \bibinfo {author} {\bibfnamefont {D.}~\bibnamefont {{Hartmann}}},\
  and\ \bibinfo {author} {\bibfnamefont {P.}~\bibnamefont {{Thaddeus}}},\
  }\bibfield  {title} {\bibinfo {title} {{The Milky Way in Molecular Clouds: A
  New Complete CO Survey}},\ }\href {https://doi.org/10.1086/318388} {\bibfield
   {journal} {\bibinfo  {journal} {\apj}\ }\textbf {\bibinfo {volume} {547}},\
  \bibinfo {pages} {792} (\bibinfo {year} {2001})},\ \Eprint
  {https://arxiv.org/abs/astro-ph/0009217} {arXiv:astro-ph/0009217 [astro-ph]}
  \BibitemShut {NoStop}%
\bibitem [{\citenamefont {Braiding}\ \emph {et~al.}(2018)\citenamefont
  {Braiding}, \citenamefont {Wong}, \citenamefont {Maxted}, \citenamefont
  {Romano}, \citenamefont {Burton}, \citenamefont {Blackwell}, \citenamefont
  {Filipović}, \citenamefont {Freeman}, \citenamefont {Indermuehle},\ and\
  \citenamefont {Lau}}]{Braiding_2018}%
  \BibitemOpen
  \bibfield  {author} {\bibinfo {author} {\bibfnamefont {C.}~\bibnamefont
  {Braiding}}, \bibinfo {author} {\bibfnamefont {G.~F.}\ \bibnamefont {Wong}},
  \bibinfo {author} {\bibfnamefont {N.~I.}\ \bibnamefont {Maxted}}, \bibinfo
  {author} {\bibfnamefont {D.}~\bibnamefont {Romano}}, \bibinfo {author}
  {\bibfnamefont {M.~G.}\ \bibnamefont {Burton}}, \bibinfo {author}
  {\bibfnamefont {R.}~\bibnamefont {Blackwell}}, \bibinfo {author}
  {\bibfnamefont {M.~D.}\ \bibnamefont {Filipović}}, \bibinfo {author}
  {\bibfnamefont {M.~S.~R.}\ \bibnamefont {Freeman}}, \bibinfo {author}
  {\bibfnamefont {B.}~\bibnamefont {Indermuehle}},\ and\ \bibinfo {author}
  {\bibfnamefont {J.}~\bibnamefont {Lau}, \bibfnamefont {{\it et al}}},\
  }\bibfield  {title} {\bibinfo {title} {The mopra southern galactic plane co
  survey—data release 3},\ }\bibfield  {journal} {\bibinfo  {journal}
  {Publications of the Astronomical Society of Australia}\ }\textbf {\bibinfo
  {volume} {35}},\ \href {https://doi.org/10.1017/pasa.2018.18}
  {10.1017/pasa.2018.18} (\bibinfo {year} {2018})\BibitemShut {NoStop}%
\bibitem [{\citenamefont {Ackermann}\ \emph {et~al.}(2012)\citenamefont
  {Ackermann}, \citenamefont {Ajello}, \citenamefont {Atwood}, \citenamefont
  {Baldini}, \citenamefont {Ballet}, \citenamefont {Barbiellini}, \citenamefont
  {Bastieri}, \citenamefont {Bechtol}, \citenamefont {Bellazzini},
  \citenamefont {Berenji} \emph {et~al.}}]{ackermann2012fermi}%
  \BibitemOpen
  \bibfield  {author} {\bibinfo {author} {\bibfnamefont {M.}~\bibnamefont
  {Ackermann}}, \bibinfo {author} {\bibfnamefont {M.}~\bibnamefont {Ajello}},
  \bibinfo {author} {\bibfnamefont {W.}~\bibnamefont {Atwood}}, \bibinfo
  {author} {\bibfnamefont {L.}~\bibnamefont {Baldini}}, \bibinfo {author}
  {\bibfnamefont {J.}~\bibnamefont {Ballet}}, \bibinfo {author} {\bibfnamefont
  {G.}~\bibnamefont {Barbiellini}}, \bibinfo {author} {\bibfnamefont
  {D.}~\bibnamefont {Bastieri}}, \bibinfo {author} {\bibfnamefont
  {K.}~\bibnamefont {Bechtol}}, \bibinfo {author} {\bibfnamefont
  {R.}~\bibnamefont {Bellazzini}}, \bibinfo {author} {\bibfnamefont
  {B.}~\bibnamefont {Berenji}}, \emph {et~al.},\ }\bibfield  {title} {\bibinfo
  {title} {Fermi-lat observations of the diffuse $\gamma$-ray emission:
  implications for cosmic rays and the interstellar medium},\ }\href@noop {}
  {\bibfield  {journal} {\bibinfo  {journal} {The Astrophysical Journal}\
  }\textbf {\bibinfo {volume} {750}},\ \bibinfo {pages} {3} (\bibinfo {year}
  {2012})}\BibitemShut {NoStop}%
\bibitem [{\citenamefont {{Wilson}}\ \emph {et~al.}(2009)\citenamefont
  {{Wilson}}, \citenamefont {{Rohlfs}},\ and\ \citenamefont
  {{H{\"u}ttemeister}}}]{2009tra..book.....W}%
  \BibitemOpen
  \bibfield  {author} {\bibinfo {author} {\bibfnamefont {T.~L.}\ \bibnamefont
  {{Wilson}}}, \bibinfo {author} {\bibfnamefont {K.}~\bibnamefont {{Rohlfs}}},\
  and\ \bibinfo {author} {\bibfnamefont {S.}~\bibnamefont
  {{H{\"u}ttemeister}}},\ }\href {https://doi.org/10.1007/978-3-540-85122-6}
  {\emph {\bibinfo {title} {{Tools of Radio Astronomy}}}}\ (\bibinfo {year}
  {2009})\BibitemShut {NoStop}%
\bibitem [{\citenamefont {{Pineda}}\ \emph {et~al.}(2008)\citenamefont
  {{Pineda}}, \citenamefont {{Caselli}},\ and\ \citenamefont
  {{Goodman}}}]{2008ApJ...679..481P}%
  \BibitemOpen
  \bibfield  {author} {\bibinfo {author} {\bibfnamefont {J.~E.}\ \bibnamefont
  {{Pineda}}}, \bibinfo {author} {\bibfnamefont {P.}~\bibnamefont
  {{Caselli}}},\ and\ \bibinfo {author} {\bibfnamefont {A.~A.}\ \bibnamefont
  {{Goodman}}},\ }\bibfield  {title} {\bibinfo {title} {{CO Isotopologues in
  the Perseus Molecular Cloud Complex: the X-factor and Regional Variations}},\
  }\href {https://doi.org/10.1086/586883} {\bibfield  {journal} {\bibinfo
  {journal} {\apj}\ }\textbf {\bibinfo {volume} {679}},\ \bibinfo {pages} {481}
  (\bibinfo {year} {2008})},\ \Eprint {https://arxiv.org/abs/0802.0708}
  {arXiv:0802.0708 [astro-ph]} \BibitemShut {NoStop}%
\bibitem [{\citenamefont {Moskalenko}\ and\ \citenamefont
  {Strong}(1998)}]{Moskalenko:1997gh}%
  \BibitemOpen
  \bibfield  {author} {\bibinfo {author} {\bibfnamefont {I.~V.}\ \bibnamefont
  {Moskalenko}}\ and\ \bibinfo {author} {\bibfnamefont {A.~W.}\ \bibnamefont
  {Strong}},\ }\bibfield  {title} {\bibinfo {title} {{Production and
  propagation of cosmic ray positrons and electrons}},\ }\href
  {https://doi.org/10.1086/305152} {\bibfield  {journal} {\bibinfo  {journal}
  {Astrophys. J.}\ }\textbf {\bibinfo {volume} {493}},\ \bibinfo {pages} {694}
  (\bibinfo {year} {1998})},\ \Eprint {https://arxiv.org/abs/astro-ph/9710124}
  {arXiv:astro-ph/9710124 [astro-ph]} \BibitemShut {NoStop}%
%%CITATION = ASTRO-PH/9710124;%%
\bibitem [{\citenamefont {Moskalenko}\ and\ \citenamefont
  {Strong}(2000)}]{Moskalenko:1998gw}%
  \BibitemOpen
  \bibfield  {author} {\bibinfo {author} {\bibfnamefont {I.~V.}\ \bibnamefont
  {Moskalenko}}\ and\ \bibinfo {author} {\bibfnamefont {A.~W.}\ \bibnamefont
  {Strong}},\ }\bibfield  {title} {\bibinfo {title} {Anisotropic inverse
  compton scattering in the galaxy},\ }\href {https://doi.org/10.1086/308138}
  {\bibfield  {journal} {\bibinfo  {journal} {Astrophys. J.}\ }\textbf
  {\bibinfo {volume} {528}},\ \bibinfo {pages} {357} (\bibinfo {year}
  {2000})},\ \Eprint {https://arxiv.org/abs/astro-ph/9811284}
  {arXiv:astro-ph/9811284 [astro-ph]} \BibitemShut {NoStop}%
%%CITATION = ASTRO-PH/9811284;%%
\bibitem [{\citenamefont {Strong}\ and\ \citenamefont
  {Moskalenko}(1998)}]{Strong:1998pw}%
  \BibitemOpen
  \bibfield  {author} {\bibinfo {author} {\bibfnamefont {A.~W.}\ \bibnamefont
  {Strong}}\ and\ \bibinfo {author} {\bibfnamefont {I.~V.}\ \bibnamefont
  {Moskalenko}},\ }\bibfield  {title} {\bibinfo {title} {{Propagation of
  cosmic-ray nucleons in the galaxy}},\ }\href {https://doi.org/10.1086/306470}
  {\bibfield  {journal} {\bibinfo  {journal} {Astrophys. J.}\ }\textbf
  {\bibinfo {volume} {509}},\ \bibinfo {pages} {212} (\bibinfo {year}
  {1998})},\ \Eprint {https://arxiv.org/abs/astro-ph/9807150}
  {arXiv:astro-ph/9807150 [astro-ph]} \BibitemShut {NoStop}%
%%CITATION = ASTRO-PH/9807150;%%
\bibitem [{\citenamefont {Strong}\ \emph {et~al.}(2000)\citenamefont {Strong},
  \citenamefont {Moskalenko},\ and\ \citenamefont {Reimer}}]{Strong:1998fr}%
  \BibitemOpen
  \bibfield  {author} {\bibinfo {author} {\bibfnamefont {A.~W.}\ \bibnamefont
  {Strong}}, \bibinfo {author} {\bibfnamefont {I.~V.}\ \bibnamefont
  {Moskalenko}},\ and\ \bibinfo {author} {\bibfnamefont {O.}~\bibnamefont
  {Reimer}},\ }\bibfield  {title} {\bibinfo {title} {{Diffuse continuum
  gamma-rays from the galaxy}},\ }\href {https://doi.org/10.1086/309038}
  {\bibfield  {journal} {\bibinfo  {journal} {Astrophys. J.}\ }\textbf
  {\bibinfo {volume} {537}},\ \bibinfo {pages} {763} (\bibinfo {year}
  {2000})},\ \bibinfo {note} {[Erratum: ApJ 541,1109(2000)]},\ \Eprint
  {https://arxiv.org/abs/astro-ph/9811296} {arXiv:astro-ph/9811296 [astro-ph]}
  \BibitemShut {NoStop}%
%%CITATION = ASTRO-PH/9811296;%%
\bibitem [{\citenamefont {{Ptuskin}}\ \emph {et~al.}(2006)\citenamefont
  {{Ptuskin}}, \citenamefont {{Moskalenko}}, \citenamefont {{Jones}},
  \citenamefont {{Strong}},\ and\ \citenamefont
  {{Zirakashvili}}}]{2006ApJ...642..902P}%
  \BibitemOpen
  \bibfield  {author} {\bibinfo {author} {\bibfnamefont {V.~S.}\ \bibnamefont
  {{Ptuskin}}}, \bibinfo {author} {\bibfnamefont {I.~V.}\ \bibnamefont
  {{Moskalenko}}}, \bibinfo {author} {\bibfnamefont {F.~C.}\ \bibnamefont
  {{Jones}}}, \bibinfo {author} {\bibfnamefont {A.~W.}\ \bibnamefont
  {{Strong}}},\ and\ \bibinfo {author} {\bibfnamefont {V.~N.}\ \bibnamefont
  {{Zirakashvili}}},\ }\bibfield  {title} {\bibinfo {title} {{Dissipation of
  Magnetohydrodynamic Waves on Energetic Particles: Impact on Interstellar
  Turbulence and Cosmic-Ray Transport}},\ }\href
  {https://doi.org/10.1086/501117} {\bibfield  {journal} {\bibinfo  {journal}
  {Astrophys. J.}\ }\textbf {\bibinfo {volume} {642}},\ \bibinfo {pages} {902}
  (\bibinfo {year} {2006})},\ \Eprint {https://arxiv.org/abs/astro-ph/0510335}
  {astro-ph/0510335} \BibitemShut {NoStop}%
\bibitem [{\citenamefont {Strong}\ \emph {et~al.}(2007)\citenamefont {Strong},
  \citenamefont {Moskalenko},\ and\ \citenamefont {Ptuskin}}]{Strong:2007nh}%
  \BibitemOpen
  \bibfield  {author} {\bibinfo {author} {\bibfnamefont {A.~W.}\ \bibnamefont
  {Strong}}, \bibinfo {author} {\bibfnamefont {I.~V.}\ \bibnamefont
  {Moskalenko}},\ and\ \bibinfo {author} {\bibfnamefont {V.~S.}\ \bibnamefont
  {Ptuskin}},\ }\bibfield  {title} {\bibinfo {title} {{Cosmic-ray propagation
  and interactions in the Galaxy}},\ }\href
  {https://doi.org/10.1146/annurev.nucl.57.090506.123011} {\bibfield  {journal}
  {\bibinfo  {journal} {ARNPS}\ }\textbf {\bibinfo {volume} {57}},\ \bibinfo
  {pages} {285} (\bibinfo {year} {2007})},\ \Eprint
  {https://arxiv.org/abs/astro-ph/0701517} {arXiv:astro-ph/0701517 [astro-ph]}
  \BibitemShut {NoStop}%
%%CITATION = ASTRO-PH/0701517;%%
\bibitem [{\citenamefont {Vladimirov}\ \emph {et~al.}(2011)\citenamefont
  {Vladimirov}, \citenamefont {Digel}, \citenamefont {Johannesson},
  \citenamefont {Michelson}, \citenamefont {Moskalenko}, \citenamefont {Nolan},
  \citenamefont {Orlando}, \citenamefont {Porter},\ and\ \citenamefont
  {Strong}}]{Vladimirov:2010aq}%
  \BibitemOpen
  \bibfield  {author} {\bibinfo {author} {\bibfnamefont {A.~E.}\ \bibnamefont
  {Vladimirov}}, \bibinfo {author} {\bibfnamefont {S.~W.}\ \bibnamefont
  {Digel}}, \bibinfo {author} {\bibfnamefont {G.}~\bibnamefont {Johannesson}},
  \bibinfo {author} {\bibfnamefont {P.~F.}\ \bibnamefont {Michelson}}, \bibinfo
  {author} {\bibfnamefont {I.~V.}\ \bibnamefont {Moskalenko}}, \bibinfo
  {author} {\bibfnamefont {P.~L.}\ \bibnamefont {Nolan}}, \bibinfo {author}
  {\bibfnamefont {E.}~\bibnamefont {Orlando}}, \bibinfo {author} {\bibfnamefont
  {T.~A.}\ \bibnamefont {Porter}},\ and\ \bibinfo {author} {\bibfnamefont
  {A.~W.}\ \bibnamefont {Strong}},\ }\bibfield  {title} {\bibinfo {title}
  {{GALPROP WebRun: an internet-based service for calculating galactic cosmic
  ray propagation and associated photon emissions}},\ }\href
  {https://doi.org/10.1016/j.cpc.2011.01.017} {\bibfield  {journal} {\bibinfo
  {journal} {Comput. Phys. Commun.}\ }\textbf {\bibinfo {volume} {182}},\
  \bibinfo {pages} {1156} (\bibinfo {year} {2011})},\ \Eprint
  {https://arxiv.org/abs/1008.3642} {arXiv:1008.3642 [astro-ph.HE]}
  \BibitemShut {NoStop}%
%%CITATION = ARXIV:1008.3642;%%
\bibitem [{\citenamefont {J\'ohannesson}\ \emph {et~al.}(2016)\citenamefont
  {J\'ohannesson} \emph {et~al.}}]{Johannesson:2016rlh}%
  \BibitemOpen
  \bibfield  {author} {\bibinfo {author} {\bibfnamefont {G.}~\bibnamefont
  {J\'ohannesson}} \emph {et~al.},\ }\bibfield  {title} {\bibinfo {title}
  {{Bayesian analysis of cosmic-ray propagation: evidence against homogeneous
  diffusion}},\ }\href {https://doi.org/10.3847/0004-637X/824/1/16} {\bibfield
  {journal} {\bibinfo  {journal} {Astrophys. J.}\ }\textbf {\bibinfo {volume}
  {824}},\ \bibinfo {pages} {16} (\bibinfo {year} {2016})},\ \Eprint
  {https://arxiv.org/abs/1602.02243} {arXiv:1602.02243 [astro-ph.HE]}
  \BibitemShut {NoStop}%
%%CITATION = ARXIV:1602.02243;%%
\bibitem [{\citenamefont {Porter}\ \emph {et~al.}(2017)\citenamefont {Porter},
  \citenamefont {Johannesson},\ and\ \citenamefont
  {Moskalenko}}]{porter2017high}%
  \BibitemOpen
  \bibfield  {author} {\bibinfo {author} {\bibfnamefont {T.~A.}\ \bibnamefont
  {Porter}}, \bibinfo {author} {\bibfnamefont {G.}~\bibnamefont
  {Johannesson}},\ and\ \bibinfo {author} {\bibfnamefont {I.~V.}\ \bibnamefont
  {Moskalenko}},\ }\bibfield  {title} {\bibinfo {title} {High-energy gamma rays
  from the milky way: Three-dimensional spatial models for the cosmic-ray and
  radiation field densities in the interstellar medium},\ }\href@noop {}
  {\bibfield  {journal} {\bibinfo  {journal} {Astrophys. J.}\ }\textbf
  {\bibinfo {volume} {846}},\ \bibinfo {pages} {23pp} (\bibinfo {year}
  {2017})}\BibitemShut {NoStop}%
\bibitem [{\citenamefont {Johannesson}\ \emph {et~al.}(2018)\citenamefont
  {Johannesson}, \citenamefont {Porter},\ and\ \citenamefont
  {Moskalenko}}]{Johannesson:2018bit}%
  \BibitemOpen
  \bibfield  {author} {\bibinfo {author} {\bibfnamefont {G.}~\bibnamefont
  {Johannesson}}, \bibinfo {author} {\bibfnamefont {T.~A.}\ \bibnamefont
  {Porter}},\ and\ \bibinfo {author} {\bibfnamefont {I.~V.}\ \bibnamefont
  {Moskalenko}},\ }\bibfield  {title} {\bibinfo {title} {{The Three-Dimensional
  Spatial Distribution of Interstellar Gas in the Milky Way: Implications for
  Cosmic Rays and High-Energy Gamma-Ray Emissions}},\ }\href
  {https://doi.org/10.3847/1538-4357/aab26e} {\bibfield  {journal} {\bibinfo
  {journal} {Astrophys. J.}\ }\textbf {\bibinfo {volume} {856}},\ \bibinfo
  {pages} {45} (\bibinfo {year} {2018})},\ \Eprint
  {https://arxiv.org/abs/1802.08646} {arXiv:1802.08646 [astro-ph.HE]}
  \BibitemShut {NoStop}%
%%CITATION = ARXIV:1802.08646;%%
\bibitem [{\citenamefont {G\'enolini}\ \emph {et~al.}(2018)\citenamefont
  {G\'enolini}, \citenamefont {Maurin}, \citenamefont {Moskalenko},\ and\
  \citenamefont {Unger}}]{PhysRevC.98.034611}%
  \BibitemOpen
  \bibfield  {author} {\bibinfo {author} {\bibfnamefont {Y.}~\bibnamefont
  {G\'enolini}}, \bibinfo {author} {\bibfnamefont {D.}~\bibnamefont {Maurin}},
  \bibinfo {author} {\bibfnamefont {I.~V.}\ \bibnamefont {Moskalenko}},\ and\
  \bibinfo {author} {\bibfnamefont {M.}~\bibnamefont {Unger}},\ }\bibfield
  {title} {\bibinfo {title} {Current status and desired precision of the
  isotopic production cross sections relevant to astrophysics of cosmic rays:
  Li, be, b, c, and n},\ }\href {https://doi.org/10.1103/PhysRevC.98.034611}
  {\bibfield  {journal} {\bibinfo  {journal} {PhRvC}\ }\textbf {\bibinfo
  {volume} {98}},\ \bibinfo {pages} {034611} (\bibinfo {year}
  {2018})}\BibitemShut {NoStop}%
\bibitem [{\citenamefont {Karwin}\ \emph {et~al.}(2019)\citenamefont {Karwin},
  \citenamefont {Murgia}, \citenamefont {Campbell},\ and\ \citenamefont
  {Moskalenko}}]{karwin2019fermi}%
  \BibitemOpen
  \bibfield  {author} {\bibinfo {author} {\bibfnamefont {C.~M.}\ \bibnamefont
  {Karwin}}, \bibinfo {author} {\bibfnamefont {S.}~\bibnamefont {Murgia}},
  \bibinfo {author} {\bibfnamefont {S.}~\bibnamefont {Campbell}},\ and\
  \bibinfo {author} {\bibfnamefont {I.~V.}\ \bibnamefont {Moskalenko}},\
  }\bibfield  {title} {\bibinfo {title} {Fermi-lat observations of $\gamma$-ray
  emission toward the outer halo of m31},\ }\href@noop {} {\bibfield  {journal}
  {\bibinfo  {journal} {The Astrophysical Journal}\ }\textbf {\bibinfo {volume}
  {880}},\ \bibinfo {pages} {95} (\bibinfo {year} {2019})}\BibitemShut
  {NoStop}%
\bibitem [{\citenamefont {{Ballet}}\ \emph {et~al.}(2020)\citenamefont
  {{Ballet}}, \citenamefont {{Burnett}}, \citenamefont {{Digel}},\ and\
  \citenamefont {{Lott}}}]{2020arXiv200511208B}%
  \BibitemOpen
  \bibfield  {author} {\bibinfo {author} {\bibfnamefont {J.}~\bibnamefont
  {{Ballet}}}, \bibinfo {author} {\bibfnamefont {T.~H.}\ \bibnamefont
  {{Burnett}}}, \bibinfo {author} {\bibfnamefont {S.~W.}\ \bibnamefont
  {{Digel}}},\ and\ \bibinfo {author} {\bibfnamefont {B.}~\bibnamefont
  {{Lott}}},\ }\bibfield  {title} {\bibinfo {title} {{Fermi Large Area
  Telescope Fourth Source Catalog Data Release 2}},\ }\href@noop {} {\bibfield
  {journal} {\bibinfo  {journal} {arXiv e-prints}\ ,\ \bibinfo {eid}
  {arXiv:2005.11208}} (\bibinfo {year} {2020})},\ \Eprint
  {https://arxiv.org/abs/2005.11208} {arXiv:2005.11208} \BibitemShut {NoStop}%
\bibitem [{\citenamefont {Blackwell}\ \emph {et~al.}(2016)\citenamefont
  {Blackwell}, \citenamefont {Burton},\ and\ \citenamefont
  {Rowell}}]{blackwell_burton_rowell_2016}%
  \BibitemOpen
  \bibfield  {author} {\bibinfo {author} {\bibfnamefont {R.}~\bibnamefont
  {Blackwell}}, \bibinfo {author} {\bibfnamefont {M.}~\bibnamefont {Burton}},\
  and\ \bibinfo {author} {\bibfnamefont {G.}~\bibnamefont {Rowell}},\
  }\bibfield  {title} {\bibinfo {title} {Mopra central molecular zone carbon
  monoxide survey status},\ }\href {https://doi.org/10.1017/S1743921316012035}
  {\bibfield  {journal} {\bibinfo  {journal} {Proceedings of the International
  Astronomical Union}\ }\textbf {\bibinfo {volume} {11}},\ \bibinfo {pages}
  {164–165} (\bibinfo {year} {2016})}\BibitemShut {NoStop}%
\bibitem [{\citenamefont {Shmakov}\ \emph {et~al.}(2022)\citenamefont
  {Shmakov}, \citenamefont {Tavakoli}, \citenamefont {Baldi}, \citenamefont
  {Karwin}, \citenamefont {Broughton},\ and\ \citenamefont
  {Murgia}}]{companion}%
  \BibitemOpen
  \bibfield  {author} {\bibinfo {author} {\bibfnamefont {A.}~\bibnamefont
  {Shmakov}}, \bibinfo {author} {\bibfnamefont {M.}~\bibnamefont {Tavakoli}},
  \bibinfo {author} {\bibfnamefont {P.}~\bibnamefont {Baldi}}, \bibinfo
  {author} {\bibfnamefont {C.}~\bibnamefont {Karwin}}, \bibinfo {author}
  {\bibfnamefont {A.}~\bibnamefont {Broughton}},\ and\ \bibinfo {author}
  {\bibfnamefont {S.}~\bibnamefont {Murgia}},\ }\href@noop {} {\bibinfo {title}
  {Deep learning models of the discrete component of the galactic interstellar
  \gray~emission}} (\bibinfo {year} {2022})\BibitemShut {NoStop}%
\end{thebibliography}%

\end{document}